\documentclass[prd,superscriptaddress]{revtex4}
\usepackage{graphicx}

\usepackage[hyperref]{hyperref}

\begin{document}

\title{Real Compton Scattering via Color Dipoles}

\author{B.Z.~Kopeliovich}

\email{Boris.Kopeliovich@usm.cl}

\affiliation{Departamento de F\'{\i}sica, Centro de Estudios Subat\'omicos, y Centro Cient\'ifico - Tecnol\'ogico de Valpara\'iso, Universidad T\'ecnica Federico Santa Mar\'{\i}a, Casilla 110-V, Valpara\'iso, Chile}

\affiliation{Joint Institute for Nuclear Research, Dubna, Russia}

\author{Ivan~Schmidt}

\email{Ivan.Schmidt@usm.cl}

\affiliation{Departamento de F\'{\i}sica, Centro de Estudios Subat\'omicos, y Centro Cient\'ifico - Tecnol\'ogico de Valpara\'iso, Universidad T\'ecnica Federico Santa Mar\'{\i}a, Casilla 110-V, Valpara\'iso, Chile}

\author{M. Siddikov}

\email{Marat.Siddikov@usm.cl}

\affiliation{Departamento de F\'{\i}sica, Centro de Estudios Subat\'omicos, y Centro Cient\'ifico - Tecnol\'ogico de Valpara\'iso, Universidad T\'ecnica Federico Santa Mar\'{\i}a, Casilla 110-V, Valpara\'iso, Chile}

\begin{abstract}
We study photoabsorption reaction and real Compton scattering (RCS)
within the color dipole model. We rely on a photon wave function derived
in the instanton vacuum model, and on the energy dependent phenomenological
elastic dipole amplitude. Data for the photoabsorption cross section
at high energies agree with our parameter free calculations. We also
provide predictions for the differential RCS cross section. Although
no data for small angle Compton scattering are available so far, this
process can be measured in ultra-peripheral hadronic and nuclear collisions
at the LHC.
\end{abstract}
\maketitle

\section{Introduction}

Compton scattering, $\gamma+p\to\gamma+p$, and related photoabsorption
reaction, have been a subject of intensive theoretical and experimental
investigation~\cite{Mueller:1998fv,Ji:1996nm,Ji:1998pc,Radyushkin:1996nd,Radyushkin:1997ki,Radyushkin:2000uy,Ji:1998xh,Collins:1998be,Collins:1996fb,Brodsky:1994kf,Goeke:2001tz,Diehl:2000xz,Belitsky:2001ns,Diehl:2003ny,Belitsky:2005qn}.
While in the case of deeply-virtual Compton scattering (DVCS), where
the initial photon is highly-virtual, the QCD factorization is proven
\cite{Radyushkin:1997ki,Collins:1998be,Ji:1998xh} and the amplitude
can be expressed in terms of the generalized parton distributions
(GPD) \cite{Mueller:1998fv,Ji:1996nm,Ji:1998pc,Radyushkin:1996nd,Radyushkin:1997ki,Radyushkin:2000uy,Ji:1998xh,Collins:1998be,Collins:1996fb,Brodsky:1994kf,Goeke:2001tz,Diehl:2000xz,Belitsky:2001ns,Diehl:2003ny,Belitsky:2005qn},
in the case of real Compton scattering (RCS) the available theoretical
tools are rather undeveloped.

On the one hand, as it has been shown in~\cite{Kronfeld:1991kp,Brooks:2000nb},
for large momentum transfer $\Delta_{\perp}$ it is possible to factorize
the RCS amplitude~\cite{Lepage:1979zb,Lepage:1980fj} and express
it in terms of the distribution amplitudes of the proton. On the other
hand, it is possible to express the amplitude of the process via the
minus 1st-moment of GPDs at zero skewedness \cite{Radyushkin:1997ki,Diehl:1998kh,Diehl:2004cx}. 

The traditional sources of quasi-real and virtual photons, the electron
beams, with very high collisions energies are expected to be available
in near future. The new projects of LHeC~\cite{Klein:2009zz,LHeC}
and EIC~\cite{EIC,EIC2} are currently under intensive discussion.
Besides the electron beams, one can~also use beams of charged hadrons.
Provided that the transverse overlap of the colliding hadrons is small,
i.e. the transverse distance $b$ between the colliding centers is
larger than the sizes of the colliding particles, $b>R_{1}+R_{2}$,
the electromagnetic interaction between colliding particles becomes
the dominant mechanism. Such processes called ultra-peripheral collisions
(UPC) can be studied in $pp$, $pA$ and $AA$ collisions. In particular,
one can access RCS in the reaction \begin{eqnarray}
A_{1}+A_{2} & \to A_{1}+\gamma+A_{2}.\label{eq:pp2ppgamma}\end{eqnarray}
 The typical virtualities $\left\langle Q_{\gamma^{*}}^{2}\right\rangle $
of the intermediate photon $\gamma^{*}$ are controlled by the formfactors
of the colliding particles, and are small: \begin{equation}
\left\langle Q_{\gamma^{*}}^{2}\right\rangle \lesssim\frac{3}{R_{A}^{2}}\sim\frac{0.1\, GeV^{2}}{A^{2/3}}.\label{eq:Q2pp2ppgamma}\end{equation}
Thus, $\left\langle Q_{\gamma^{*}}^{2}\right\rangle $ is of the order
of the soft hadronic scale, so the intermediate photon can be treated
as a free Weizs\"acker-Williams one, i.e. the amplitude of the process~(\ref{eq:pp2ppgamma})
can be described in terms of RCS. 

These processes at the LHC will allow to study RCS at very high energies.
The possibility of observation of such processes experimentally has
been demonstrated by the STAR \cite{Abelev:2007nb,Adams:2004rz,Adler:2002sc}
and PHENIX~\cite{dEnterria:2006ep} experiments at RHIC. It is expected
that at LHC photon-proton collisions at energies up to $\sqrt{s_{\gamma p}}\lesssim8\times10^{3}$~GeV
can be observed~\cite{Baltz:2007kq}. In this paper we concentrate
on RCS on a proton target. Nuclear effects will be discussed elsewhere.

In what follows we employ the color dipole approach introduced in~\cite{Kopeliovich:1981,mueller}.
The central objects of the model are the dipole scattering amplitude,~$\mathcal{A}(s,\beta,\vec{r})$
and the light-cone quark distribution functions of the photon. While
pQCD predicts the dipole amplitude only for small-size dipoles, several
successful phenomenological parameterizations for the large-size dipoles
are known. Relying on the photon wave function evaluated in the instanton
vacuum model~\cite{Dorokhov:2006qm}, which is valid for any $Q^{2}$,
one can extend the applicability of the model to the case of the processes
with real photons~\cite{Kopeliovich:2008ct}. In this paper we are
going to consider the real photoabsorption, $\gamma+p\to X$, and
the RCS.

\section{Color dipole model}

\label{sec:DipoleModel} The color dipole model is valid only at sufficiently
high energies, where the dominant contribution to the Compton amplitude
comes from gluonic exchanges. Then the general expression for the
Compton amplitude in the color dipole model has the form,

\begin{equation}
\mathcal{A}_{\mu\nu}\left(s,\Delta\right)\approx e_{\mu}^{(i)}e_{\nu}^{(j)}\int d\beta_{1}d\beta_{2}d^{2}r_{1}d^{2}r_{2}\bar{\Psi}_{\gamma}^{(i)}\left(\beta_{2},\vec{r}_{2}\right)\mathcal{A}^{d}\left(\beta_{1},\vec{r}_{1};\beta_{2},\vec{r}_{2};\Delta\right)\Psi_{\gamma}^{(j)}\left(\beta_{1},\vec{r}_{1}\right),\label{eq:Convolution:Full}\end{equation}
 where $e_{\mu}^{(i)}$ is the photon polarization vector; $\beta_{1,2}$
are the light-cone fractional momenta of the quark and antiquark,
$\vec{r}_{1,2}$ are the transverse distances in the final and initial
dipoles respectively;\textbf{ }$\Delta$ is the momentum transfer
in the Compton scattering\textbf{,} $\mathcal{A}^{d}(...)$ is the
scattering amplitude for the dipole state which also implicitly depends
on $s$, c.m. energy squared, and $\Psi_{\gamma}^{(i)}\left(\beta_{2},\vec{r}_{2}\right)$
is the wavefunction of the photon in the polarization state $i$ \cite{Dorokhov:2006qm}.

At high energies in the small angle approximation, $\Delta/\sqrt{s}\ll1$,
the quark separation and fractional momenta are preserved, so, \begin{eqnarray}
\mathcal{I}m\,\mathcal{A}^{d}\left(\beta_{1},\vec{r}_{1};\beta_{2},\vec{r}_{2};\Delta\right) & \approx & \delta\left(\beta_{1}-\beta_{2}\right)\delta\left(\vec{r}_{1}-\vec{r}_{2}\right)Im\, f_{\bar{q}q}^{N}(\vec{r}_{1},\vec{\Delta},\beta_{1}).\label{eq:DVCSIm-BK}\end{eqnarray}
Generally, the amplitude $f_{\bar{q}q}^{N}(...)$ is a nonperturbative
object, with asymptotic behavior for small~$r$ controlled by pQCD
\cite{Kopeliovich:1981}:\[
f_{\bar{q}q}^{N}(\vec{r},\vec{\Delta},\beta)\sim r^{2},\]
 up to slowly varying corrections $\sim\ln(r)$. 

Calculation of the RCS differential cross section also involves the
real part of scattering amplitude, whose relation to the imaginary
part is quite straightforward. According to \cite{Bronzan:1974jh},
if the limit $\lim\limits _{s\to\infty}\left(\frac{\mathcal{I}m\,{f}}{s^{\alpha}}\right)$
exists and is finite, then the real part and imaginary parts of the
forward amplitude are related as\begin{equation}
\mathcal{R}e\,{f(\Delta=0)}=s^{\alpha}\tan\left[\frac{\pi}{2}\left(\alpha-1+\frac{\partial}{\partial\ln s}\right)\right]\frac{\mathcal{I}m\,{f(\Delta=0)}}{s^{\alpha}}.\label{eq:BronzanFul}\end{equation}
 In the model under consideration the imaginary part of the forward
dipole amplitude indeed has a power dependence on energy, $\mathcal{I}m\, f(\Delta=0)(s)\sim s^{\alpha}$,
so (\ref{eq:BronzanFul}) simplifies to \begin{eqnarray}
\frac{\mathcal{R}e\,\mathcal{A}}{\mathcal{I}m\,\mathcal{A}} & =\tan\left(\frac{\pi}{2}(\alpha-1)\right)\equiv\epsilon.\end{eqnarray}

This fixes the phase of the forward Compton amplitude, which we retain
for nonzero momentum transfers assuming for the real and imaginary
parts similar dependences on. Finally we arrive at, \begin{equation}
\mathcal{A}_{\mu\nu}\approx(\epsilon+i)e_{\mu}^{(i)}(q')e_{\nu}^{(j)}(q)\int d^{2}r\int d\beta\,\bar{\Psi}_{\gamma}^{(i)}(\beta,r)\Psi_{\gamma}^{(j)}(\beta,r)\, Im\, f_{\bar{q}q}^{N}(\vec{r},\vec{\Delta},\beta,s),\label{eq:DVCS-Im-conv}\end{equation}

For the cross-section of unpolarized Compton scattering, from~(\ref{eq:DVCS-Im-conv})
we obtain, \begin{eqnarray}
\frac{d\sigma_{el}^{\gamma p}}{dt} & = & \frac{1+\epsilon^{2}}{16\pi}\sum_{ij}\left|\mathcal{A}_{\mu\nu}^{(ij)}\right|^{2}=\nonumber \\
 & = & \frac{1+\epsilon^{2}}{16\pi}\sum_{ij}\left|\int d^{2}r\int d\beta\,\bar{\Psi}_{\gamma}^{(i)}(\beta,r)\Psi_{\gamma}^{(j)}(\beta,r)\, Im\, f_{\bar{q}q}^{N}(\vec{r},\vec{\Delta},\beta)\right|^{2}.\label{eq:DVCS-cross-section}\end{eqnarray}

The imaginary part of the forward amplitude~(\ref{eq:DVCS-Im-conv})
gives the total photoabsorption cross-section, \begin{equation}
\sigma_{tot}^{\gamma p}=\frac{1}{16\pi}\int d\beta d^{2}r\left|\Psi_{\gamma}(\beta,r)\right|^{2}Im\, f_{\bar{q}q}^{N}\left(\vec{r},\vec{\Delta},\beta\right).\label{eq:photo-tot}\end{equation}

Formulas~(\ref{eq:DVCS-cross-section},\ref{eq:photo-tot}) are used
further for numerical calculations.

\section{Wavefunctions from the instanton vacuum}

\label{sec:WFfromIVM}In this section we present briefly some details
 of the wavefunction evaluation in the instanton vacuum model~(see~\cite{Schafer:1996wv,Diakonov:1985eg,Diakonov:1995qy}
and references therein). The central object of the model is the partition function of the light quarks, which has the form
\begin{equation}
Z[v]=\int d\lambda\mathcal{D}\bar\psi\mathcal{D}\psi\mathcal{D}\Phi e^{iS[\lambda, v,\bar\psi,\psi,\Phi]},
\end{equation}
where the effective action $S[\lambda, v,\bar\psi,\psi,\Phi]$ is defined as~\cite{Diakonov:1995qy,Goeke:2007j}
\[
S[\lambda, v,\bar\psi,\psi,\Phi]=\int d^{4}x\left(\frac{N}{V}\ln\lambda+2\Phi^{2}(x)-\bar{\psi}\left(\hat{p}+\hat{v}-m-c\bar{L}f\otimes\Phi\cdot\Gamma_{m}\otimes fL\right)\psi\right).\]
 Here  $\psi$
and $\Phi$ are the fields of constituent quarks and mesons respectively,
$N/V$ is the density of the instanton gas, $\hat{v}\equiv v_{\mu}\gamma^{\mu}$
is the external vector current corresponding to the photon, $L$ is
the gauge factor, \begin{equation}
L\left(x,z\right)=P\exp\left(i\int_{z}^{x}d\zeta^{\mu}v_{\mu}(\zeta)\right),\label{eq:L-factor}\end{equation}
 which provides the gauge covariance of the action, and the nonlinear term in explicit form is
\begin{equation}
 -c\bar \psi\bar{L}f\otimes\Phi\cdot\Gamma_{m}\otimes fL\psi \equiv -c\int d^4 x d^4 y d^4 z \bar \psi(x)\bar{L}\left(x-z\right)\tilde f(x-z)\left(\sum_i\Phi_i(z)\Gamma_{m,i}\right) \tilde f(z-y)L(z-y)\psi(y)
\end{equation}
 where $\Gamma_{m}$ is one of the matrices, $\Gamma_{m}=\left\{1,i\vec{\tau},\gamma_{5},i\vec{\tau}\gamma_{5}\right\}$, $\tilde f(x-y)=\int \frac{d^4p}{(2\pi)^4}f(p)e^{-ip\cdot(x-y)}f(p)$, and $f(p)$ is the Fourier transform of the zero-mode profile.

In the leading order in $N_{c}$, we have the same Feynman rules as
in the perturbative theory, but with momentum-dependent quark mass
$\mu(p)$ in the quark propagator\begin{eqnarray}
S(p) & = & \frac{1}{\hat{p}-\mu(p)+i0}.\end{eqnarray}
The mass of the constituent quark has a form \[
\mu(p)=m+M\, f^{2}(p),\]
 where $m\approx5$~MeV is the current quark mass, $M\approx350$~MeV
is the dynamical mass generated by the interaction with the instanton
vacuum background. Due to the presence of  instantons the coupling
of a vector current to a quark is also modified,\begin{eqnarray}
\hat{v} & \equiv & v_{\mu}\gamma^{\mu}\rightarrow\hat{V}=\hat{v}+\hat{V}^{nonl},\nonumber \\
\hat{V}^{nonl} & \approx & -2Mf(p)\frac{df(p)}{dp_{\mu}}v_{\mu}(q)+\mathcal{O}\left(q^{2}\right),\label{eq:V-nonl-expanded}\end{eqnarray}
where $p$ is the momentum of the incoming quark, and $q$ is the momentum of the photon.
Notice that for an arbitrary photon momentum $q$ the expression for
$\hat{V}^{nonl}$ depends on the choice of the path in~(\ref{eq:L-factor})
and as a result one can find in the literature different expressions
used for evaluations~\cite{Dorokhov:2006qm,Anikin:2000rq,Dorokhov:2003kf,Goeke:2007j}.
In the limit $p\to\infty$ the function $f(p)$ falls off as $\sim\frac{1}{p^{3}},$
so for large $p\gg\rho^{-1}$, where $\rho\approx(600\, MeV)^{-1}$
is the mean instanton size, the mass of the quark $\mu(p)\approx m$
and vector current interaction vertex $\hat{V}\approx\hat{v}$. However
we would like to emphasize that the wavefunction $\Psi(\beta,r)$
gets contribution from both the soft and the hard parts, so even in
the large-$Q$ limit the instanton vacuum function is different from
the well-known perturbative result.

We have to evaluate the wavefunctions associated with the following
matrix elements:

\begin{eqnarray}
I_{\Gamma}(\beta,\vec{r}) & = & \int\frac{dz}{2\pi}e^{i\left(\beta+\frac{1}{2}\right)q\cdot z}\left\langle 0\left|\bar{\psi}\left(-\frac{z}{2}n-\frac{\vec{r}}{2}\right)\Gamma\psi\left(\frac{z}{2}n+\frac{\vec{r}}{2}\right)\right|\gamma(q)\right\rangle ,\end{eqnarray}
where $\Gamma$ is one of the matrices $\Gamma=\left\{\gamma_{\mu},\gamma_{\mu}\gamma_{5},\sigma_{\mu\nu}\right\}.$
In the leading order in $N_{c}$ one can easily obtain\begin{equation}
I_{\Gamma}=\int\frac{d^{4}p}{(2\pi)^{4}}e^{i\vec{p}_{\perp}\vec{r}_{\perp}}\delta\left(p^{+}-\left(\beta+\frac{1}{2}\right)q^{+}\right)Tr\left(S(p)\hat{V}S(p+q)\Gamma\right).\label{eq:WF-LO}\end{equation}
The evaluation of~(\ref{eq:WF-LO}) is quite tedious but straightforward.
Details of this evaluation may be found in~\cite{Dorokhov:2006qm}.

The overlap of the initial and final photon wavefunctions in~(\ref{eq:DVCS-cross-section})
was evaluated according to\begin{equation}
\Psi^{(i)*}(\beta,r,Q^{2}=0)\Psi^{(i)}(\beta,r,Q^{2})=\sum_{\Gamma}I_{\Gamma}^{*}\left(\beta,r^{*},0\right)I_{\Gamma}\left(\beta,r,Q^{2}\right),\label{eq:ConvolutionWF}\end{equation}
 where summation is done over possible polarization states $\Gamma=\left\{\gamma_{\mu},\gamma_{\mu}\gamma_{5},\sigma_{\mu\nu}\right\}$.
In the final state we should use $r_{\mu}^{*}=r_{\mu}+n_{\mu}\frac{q_{\perp}'\cdot r_{\perp}}{q_{+}}=r_{\mu}-n_{\mu}\frac{\Delta_{\perp}\cdot r_{\perp}}{q_{+}}$,
which is related to the reference frame with $q'_{\perp,\mu}=0$ in which
the components~(\ref{eq:WF-LO}) were evaluated.

\section{Numerical results}

\label{sec:Results}

\subsection{Photoabsorption}

The Bjorken variable used in DIS, $x=Q^{2}/\left(2\, p\cdot q\right)$,
is not appropriate at small photon virtualities, since it does not
have the meaning of a fractional quark momentum any more, and may
be very small even at low energies. In particular, for RCS the Bjorken
variable $x$ defined in this way would be zero. Therefore, one should
rely on the phenomenological dipole cross section which depends on
energy, rather than $x$. We use the $s$-dependent dipole cross section
proposed in \cite{kst2}, which saturates at large separations in
analogy to the $x$-dependent one proposed in \cite{gbw}. Correspondingly,
for the elastic dipole amplitude we employ the model developed in
\cite{Kopeliovich:2007fv,Kopeliovich:2008nx,Kopeliovich:2008da,Kopeliovich:2008ct},
\begin{eqnarray}
Im\, f_{\bar{q}q}^{N}(\vec{r},\vec{\Delta},\beta,s) & = & \frac{\sigma_{0}(s)}{4}\exp\left[-\left(\frac{B(s)}{2}+\frac{R_{0}^{2}(s)}{16}\right)\vec{\Delta}_{\perp}^{2}\right]\times\nonumber \\
 & \times & \left(e^{-i\beta\vec{r}\cdot\vec{\Delta}}+e^{i(1-\beta)\vec{r}\cdot\vec{\Delta}}-2e^{i\left(\frac{1}{2}-\beta\right)\vec{r}\cdot\vec{\Delta}}e^{-\frac{r^{2}}{R_{0}^{2}(s)}}\right),\label{eq:Im_f-result}\end{eqnarray}
 where $\sigma_{0}(s),\, R_{0}^{2}(s),\, B(s)$ are the phenomenological
parameters known from DIS and $\pi p$ scattering data.

We employ the $s$-dependent parametrization of the dipole cross section
suitable for soft processes \cite{kst2} \begin{eqnarray}
\sigma_{\bar{q}q}(r,s) & = & \sigma_{0}(s)\left(1-e^{-r^{2}/R_{0}^{2}(s)}\right),\label{eq:dipole-1}\\
\sigma_{0}(s) & = & \sigma_{\pi p}(s)\left(1+\frac{3}{8}\frac{R_{0}^{2}(s)}{r_{\pi}^{2}}\right),\label{eq:dipole-2}\\
R_{0}(s) & = & 0.88\, fm\times\left(\frac{s_{0}}{s}\right)^{0.14},\label{eq:R0_standard}\end{eqnarray}
 where $s_{0}\approx1000\, GeV^{2}.$ For the pion cross section we
use the parametrization and fit of~\cite{Donnachie:1992ny}, namely
its Pomeron part, \begin{equation}
\sigma_{\pi p}(s)=23.6\left(\frac{s}{s_{0}}\right)^{0.079}mb.\label{eq:dipole-3}\end{equation}
The parameter $B(s)$ in Eq.~(\ref{eq:Im_f-result}), is related
to the $t$-slope of the differential cross section of elastic $\pi p$
scattering \cite{Kopeliovich:2007fv,Kopeliovich:2008da,Kopeliovich:2008nx},
\begin{equation}
B(s)=B_{el}^{\pi p}(s)-\frac{1}{3}\left\langle r_{ch}^{2}\right\rangle _{\pi}-\frac{1}{8}\, R_{0}^{2}(s)\label{eq:B-slope}\end{equation}
Here we rely on the Regge factorization and use $B_{el}^{\pi p}(s)=B_{0}+2\alpha_{P}'(0)\ln\left(s/\mu_{0}^{2}\right)$,
with $B_{0}=6$~GeV$^{-2},\left\langle r_{ch}^{2}\right\rangle _{\pi}=0.44$~fm$^{2}$,$\alpha'_{P}(0)=0.25$~GeV$^{-2}$,
and $\mu_{0}=1$~GeV.

\begin{figure}[ht]
\includegraphics[scale=0.4]{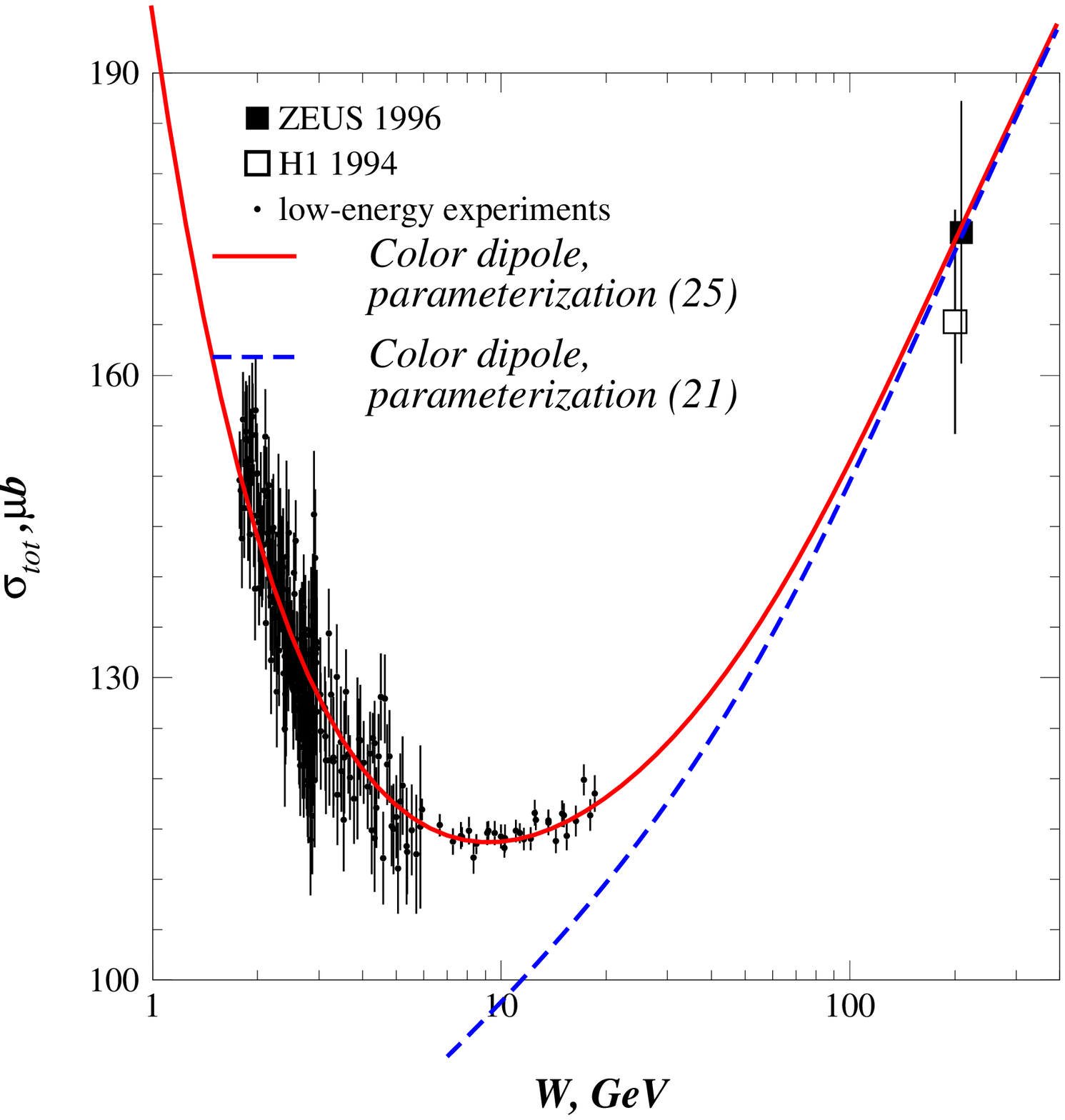}
\includegraphics[scale=0.4]{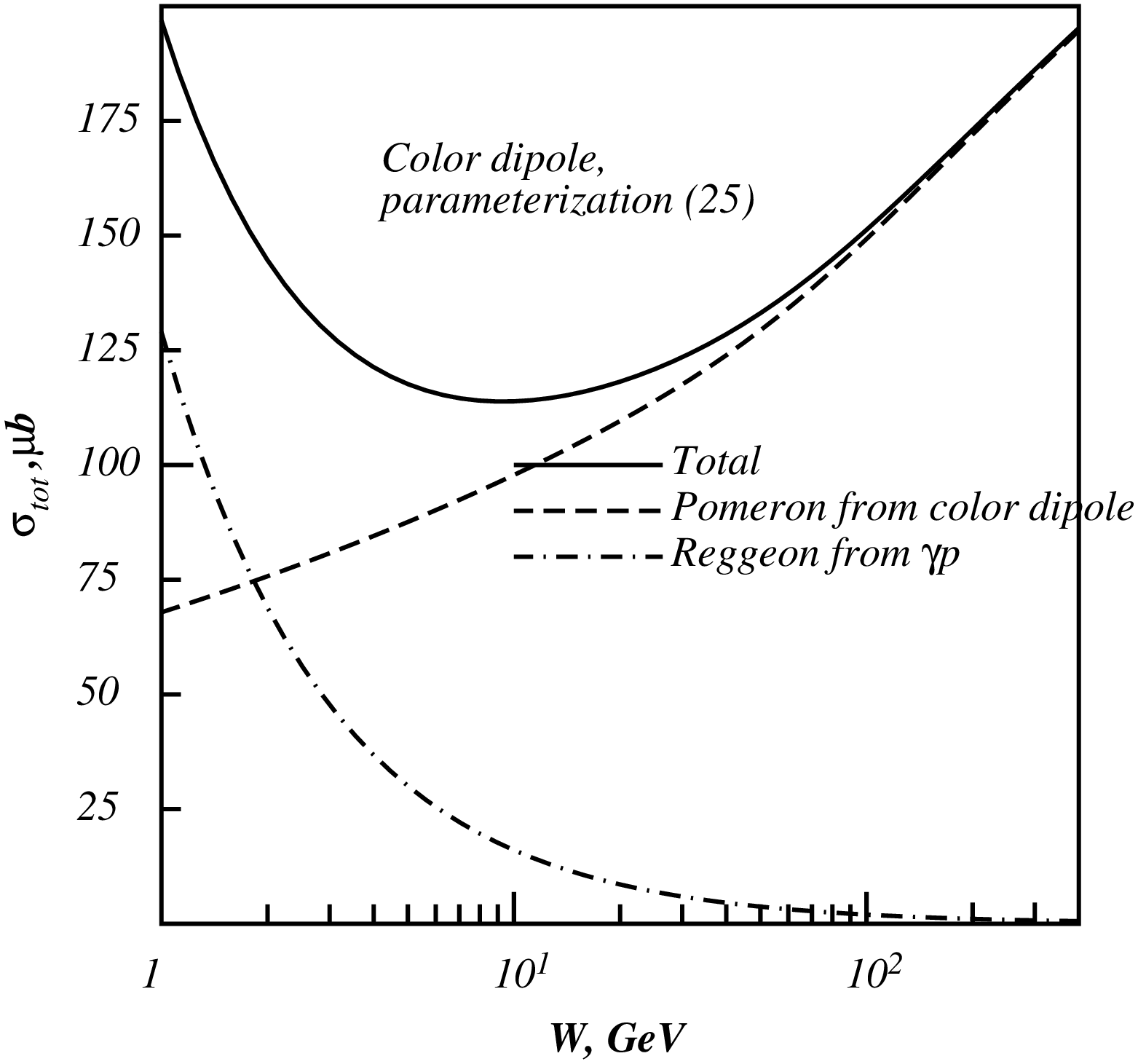}

\caption{\label{fig:PhAXSection} [Color online] Photoabsorption cross-section
in the color dipole model as function of c.m. energy, $W=\sqrt{s}$.
Left: comparison of calculations with experimental data from ZEUS~\cite{Chekanov:2001gw}.
Dashed line corresponds to the parameterization~(\ref{eq:dipole-1}-\ref{eq:R0_standard}),
solid line corresponds to addition of the Reggeon term,Eq.~(\ref{eq:R0_modified}).
Right: the contributions of the Pomeron and Reggeon parts plotted
separately. At $W\gtrsim10$~GeV the Reggeon contribution becomes
negligibly small.}

\end{figure}

This parametrization may be used in Eqs.~(\ref{eq:DVCS-cross-section}-\ref{eq:photo-tot})
only at very high energies where in terms of the Regge theory, the
Pomeron term in the cross section dominates. So far only two data
points shown in Fig.~\ref{fig:PhAXSection} are available for $\sigma_{\gamma p}$
from the H1 and ZEUS experiments \cite{Chekanov:2001gw}, and our
parameter free calculation agrees well with these data.

In order to extend the model down to smaller values of $\sqrt{s}$,
where more data are available, we added the Reggeon contribution,
which was fitted to the photoabsorption data in \cite{Donnachie:1992ny}
\begin{equation}
\sigma_{\gamma p}^{(R)}(s)=129\, nb\times s^{-0.4525}.\label{eq:reggeon-photo}\end{equation}

Besides, the Pomeron part of the dipole cross section parametrized
as in Eqs.~(\ref{eq:dipole-1}-\ref{eq:R0_standard}) exposes some
problems at low energies. Indeed, as one can see from~(\ref{eq:R0_standard}),
the saturation radius $R_{0}(s)$ grows and may substantially exceed
the confinement radius. In order to regularize the low-energy behavior
of $R_{0}(s)$, we modify Eq.~(\ref{eq:R0_standard}) as follows,
\begin{eqnarray}
R_{0}(s)\Rightarrow\tilde{R}_{0}(s)=0.88\, fm\left(\frac{s_{0}}{s+s_{1}}\right)^{0.14}.\label{eq:R0_modified}\end{eqnarray}

Fit to low-energy photoabsorption data allows to fix this parameter
at $\sqrt{s}_{1}=60$~GeV ~(see Fig.~(\ref{fig:PhAXSection}) for
more details). Since further evaluations are done in the LHC energy
range, the difference between~(\ref{eq:R0_standard}) and (\ref{eq:R0_modified})
is negligible, as one can see from Fig.~(\ref{fig:PhAXSection}).
Indeed, both parameterizations coincide for $\sqrt{s}\gtrsim10$ GeV.
In the right pane of the Fig.~(\ref{fig:PhAXSection}) the contributions
of the color dipole~(\ref{eq:photo-tot}) and Reggeon~(\ref{eq:reggeon-photo})
terms are displayed separately.

\subsection{Compton scattering}

\begin{figure}[ht]
\includegraphics[scale=0.4]{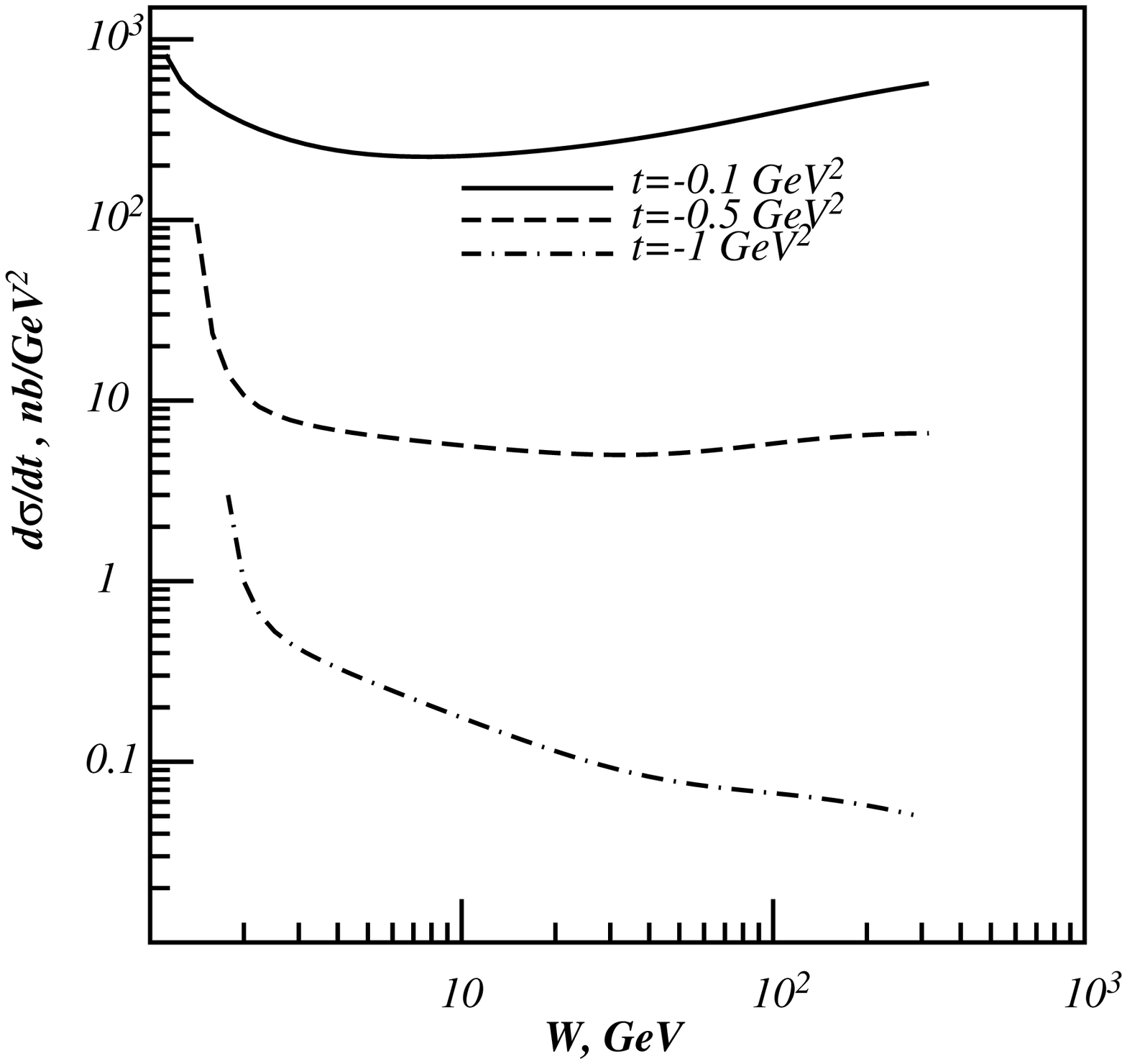}
\includegraphics[scale=0.4]{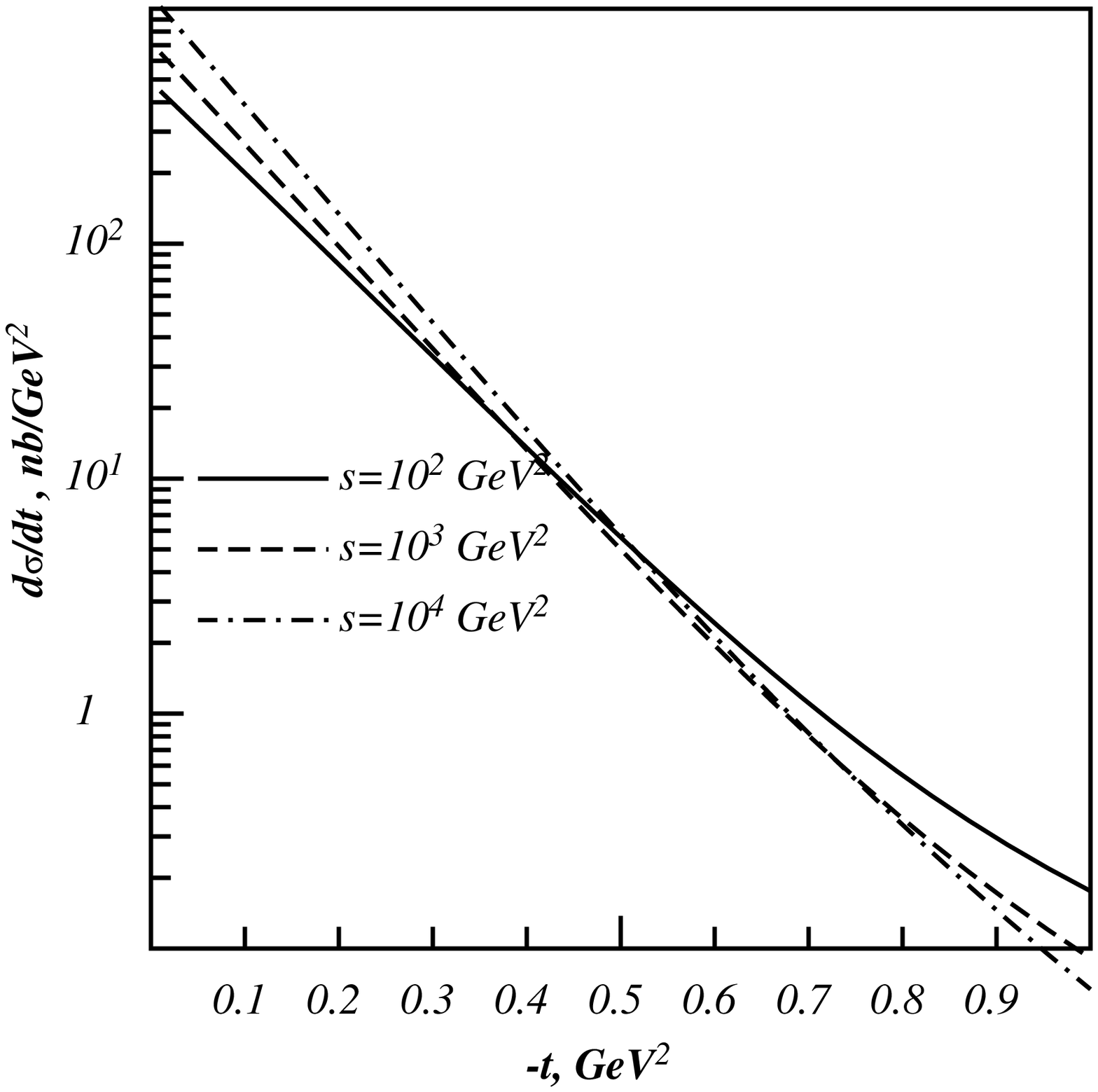}

\caption{\label{fig:RCSXSection} RCS cross-section in the color dipole model.
Left: Energy dependence of the RCS cross-section for different $t$,
$W\equiv\sqrt{s_{\gamma p}}$. For $s\gtrsim10$~GeV$^{2}$ the cross-section
depends on the energy $s$ approximately as $s^{2(\alpha(t)-1)}$.
For $s\lesssim10$~GeV$^{2}$ we have ``soft'' regime where contributions
of reggeons dominate. Right: $t$-dependence of the RCS cross-section
for different energies in the energy range of ultraperipheral collisions
at LHC . }

\end{figure}

Using parameterization~(\ref{eq:dipole-1}-\ref{eq:dipole-3}), we
calculate the elastic RCS differential cross-section as
\begin{eqnarray}
\frac{d\sigma_{el}^{\gamma p}}{dt}
 & = & \frac{1+\epsilon^{2}}{16\pi}\sum_{ij}\left|\int d^{2}r\int d\beta\,\bar{\Psi}_{\gamma}^{(i)}(\beta,r)\Psi_{\gamma}^{(j)}(\beta,r)\, Im\, f_{\bar{q}q}^{N}(\vec{r},\vec{\Delta},\beta)+16\pi\sigma_{\gamma p}^{(I\!\!R)}(s)e^{B_{I\!\!R}(s)t}\right|^{2}.\label{eq:DVCS-cross-section-full}\end{eqnarray}
Since extraction of reggeon parameters from experimental data yields huge uncertainties~\cite{Aktas:2006hx}, in this paper we rely on the $f$-dominance of the pomeron~\cite{Irving:1976rk} and take the reggeon slope as
\begin{equation}
B_{I\!\!R}(s)=B_1+2\alpha'_{I\!\!R}(0)\ln\left(\frac{s}{\mu_0^2}\right),\\
\end{equation}
 where $B_1=B_0=6$ GeV$^{-2}$, $\alpha'_{I\!\!R}(0)=0.9$ GeV$^{-2}$,
and the results for the differential cross-section
are presented in the Figure~\ref{fig:RCSXSection}. As one can see
from the left pane, for $s\gtrsim10$~GeV$^{2}$ the cross-section
rises with energy for small $|t|$, but falls at $|t|=1$GeV$^{2}$.
This corresponds to the Regge predicted energy dependence $s^{2(\alpha(t)-1)}$.
However, a word of caution is in order here, since the linear $t$-dependence
of the Pomeron trajectory may not continue at large $|t|$, and indeed
data indicate that $\alpha_{P}(t)$ levels off~\cite{Brandt:1997gi}.
On the right pane of the Figure~\ref{fig:RCSXSection} our predictions
for the $t$-dependence of the cross section are plotted for different
energies. The cross section demonstrates a shrinkage of diffraction
cone with energy in accordance with the Regge theory. 

\begin{figure}[ht]
\includegraphics[scale=0.4]{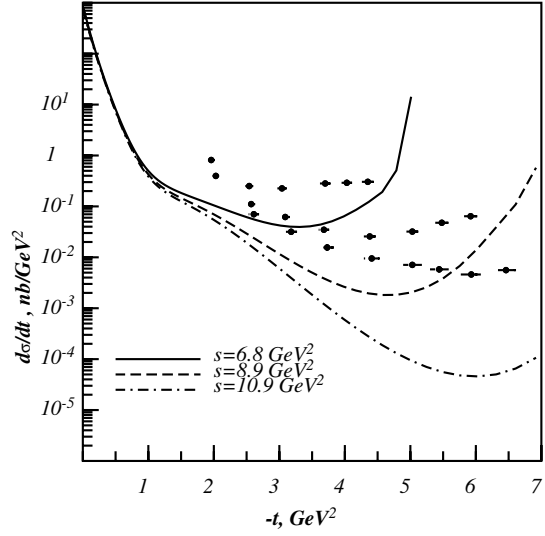}

\caption{\label{fig:RCS-JLAB}Comparison of the RCS cross-section evaluated
in the color dipole model with low-energy (large-angle) experimental
data from JLAB \cite{Danagoulian:2007gs}.}

\end{figure}

The RCS cross-section has been measured so far only at Jefferson Lab
(JLAB) at energies $s\lesssim10\, GeV^{2}$~\cite{Danagoulian:2007gs}.
In Figure~\ref{fig:RCS-JLAB} we compare predictions of the color
dipole model with experimental data. Since these data also have relatively
large $|t|\gtrsim$2 GeV$^{2}$ (wide-angle Compton scattering), calculations
in the dipole approach go beyond the kinematics of validity of the
model. Indeed, Eqn.~(\ref{eq:Im_f-result}), lead to the RCS cross-section
which decreases exponentially at fixed $s/t$, while the general pQCD
analysis~\cite{Brodsky:1973kr} predicts $1/s^{6}$ behavior.

As was discussed in the introduction, there are two approaches which
are used to describe the wide-angle Compton scattering. The first
one is valid for large~$\Delta_{\perp}$\cite{Lepage:1979zb,Lepage:1980fj}
and expresses the amplitude via the distribution amplitude of three
valence quarks in the proton. The RCS cross-section in this approach
was studied in~\cite{Kronfeld:1991kp,Brooks:2000nb} and it was
found that evaluation with widely used distribution amplitudes also
underestimates the JLAB data~\cite{Danagoulian:2007gs}. Another
description expresses the RCS amplitude via the $1/x$ moment of GPDs
at zero skewedness,~$\int\frac{dx}{x}H(x,0,t)$~\cite{Radyushkin:1997ki,Diehl:1998kh,Diehl:2004cx}.
This approach is able to describe the existing JLAB data. However
the $t$-dependence of the cross-section in this approach
depends on the model for the GPD used in the evaluation.

\begin{figure}[ht]
\includegraphics[scale=0.5,angle=-90]{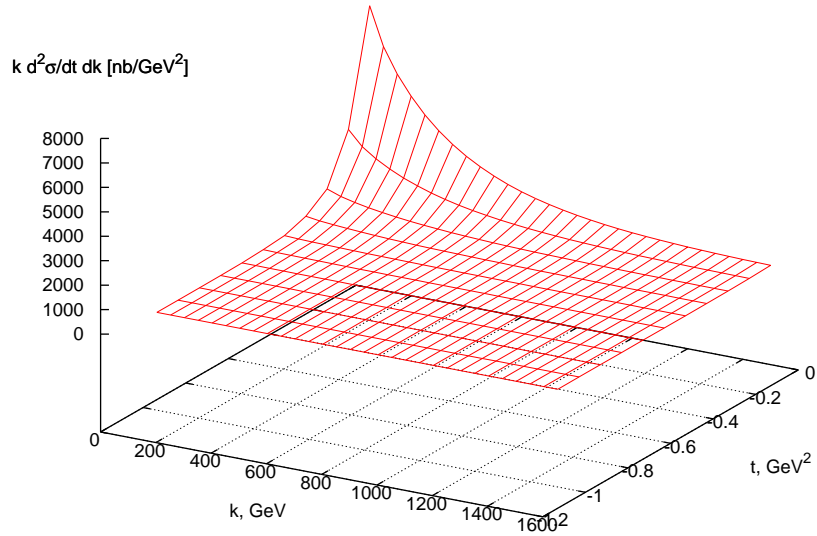}
\caption{\label{fig:TotalXSection}[Color online] Twofold cross-section $k d^2\sigma/dt dk$}
\end{figure}

Since the experimental counting rate also includes the flux of quasi-real photons, we present in Fig.~\ref{fig:TotalXSection} the two-dimensional product of the flux and the differential RCS cross section,
\begin{equation}
\frac{d^2 \sigma_{pp\to pp\gamma}}{dk dt}=\frac{dN_\gamma}{dk} \frac{d\sigma_{\gamma p\to \gamma p}}{dt},
\end{equation}
where $k$ is the absolute value of the wave vector of the quasireal photon, and photon flux $\frac{dN_\gamma}{dk}$ is given, e.g., in~\cite{Baltz:2007kq}.

\section{Summary}

\label{sec:Conclusions} In this paper we evaluated the photoabsorption
and RCS cross-sections within the color dipole model. We employed
a photon wave function calculated in the instanton-vacuum model and
incorporating nonperturbative effects~\cite{Dorokhov:2006qm}. For
the dipole cross section we relied on an energy dependent parametrization,
because Bjorken $x$ is not a proper variable at low photon virtualities.

We found that the model describes available data for the photoabsorption
cross section from the ZEUS and H1 data quite well, justifying application
of the color dipole for processes with real photons. Applicability
of the dipole model was extended down to smaller energies, $\sqrt{s}\lesssim30$
GeV, by freezing the saturation radius $R_{s}(s)$ Eq.~(\ref{eq:R0_standard})
for the energy dependent parametrization~\cite{kst2}, to make sure
that it does not exceed the confinement radius.

We also evaluated the RCS cross-sections and made predictions for
the energy range to be accessed in ultra-peripheral collisions at
LHC (Figure~\ref{fig:RCSXSection}).

\section*{Acknowlegments}

This work was supported in part by Fondecyt (Chile) grants 1090291,
1090073 , and by DFG (Germany) grant PI182/3-1.


\begin{thebibliography}{10}
 \bibitem{Mueller:1998fv} D.~Mueller, D.~Robaschik, B.~Geyer, F.~M.~Dittes
 and J.~Horejsi, Fortsch.\ Phys.\ \textbf{42}, 101 (1994) [arXiv:hep-ph/9812448].
 
 \bibitem{Ji:1996nm} X.~D.~Ji, Phys.\ Rev.\ D \textbf{55}, 7114
 (1997).
 
 \bibitem{Ji:1998pc} X.~D.~Ji, J.\ Phys.\ G \textbf{24}, 1181
 (1998) [arXiv:hep-ph/9807358].
 
 \bibitem{Radyushkin:1996nd} A.~V.~Radyushkin, Phys.\ Lett.\ B
 \textbf{380}, 417 (1996) [arXiv:hep-ph/9604317].
 
 \bibitem{Radyushkin:1997ki} A.~V.~Radyushkin, Phys.\ Rev.\ D
 \textbf{56}, 5524 (1997).
 
 \bibitem{Radyushkin:2000uy} A.~V.~Radyushkin, arXiv:hep-ph/0101225.
 
 \bibitem{Ji:1998xh} X.~D.~Ji and J.~Osborne, Phys.\ Rev.\ D
 \textbf{58} (1998) 094018 [arXiv:hep-ph/9801260].
 
 \bibitem{Collins:1998be}J.~C.~Collins and A.~Freund,   Phys.~Rev.~D \textbf{59} (1999) 074009 [arXiv:hep-ph/9801262].
 
 \bibitem{Collins:1996fb} J.~C.~Collins, L.~Frankfurt and M.~Strikman,
 Phys.\ Rev.\ D \textbf{56}, 2982 (1997).
 
 \bibitem{Brodsky:1994kf} S.~J.~Brodsky, L.~Frankfurt, J.~F.~Gunion,
 A.~H.~Mueller and M.~Strikman, Phys.\ Rev.\ D \textbf{50}, 3134
 (1994).
 
 \bibitem{Goeke:2001tz} K.~Goeke, M.~V.~Polyakov and M.~Vanderhaeghen,
 Prog.\ Part.\ Nucl.\ Phys.\ \textbf{47}, 401 (2001) [arXiv:hep-ph/0106012].
 
 \bibitem{Diehl:2000xz} M.~Diehl, T.~Feldmann, R.~Jakob and P.~Kroll,
 Nucl.\ Phys.\ B \textbf{596}, 33 (2001) [Erratum-ibid.\ B \textbf{605},
 647 (2001)] [arXiv:hep-ph/0009255].
 
 \bibitem{Belitsky:2001ns} A.~V.~Belitsky, D.~Mueller and A.~Kirchner,
 Nucl.\ Phys.\ B \textbf{629}, 323 (2002) [arXiv:hep-ph/0112108].
 
 \bibitem{Diehl:2003ny} M.~Diehl, Phys.\ Rept.\ \textbf{388}, 41
 (2003) [arXiv:hep-ph/0307382].
 
 \bibitem{Belitsky:2005qn} A.~V.~Belitsky and A.~V.~Radyushkin,
 Phys.\ Rept.\ \textbf{418}, 1 (2005) [arXiv:hep-ph/0504030].
 
 \bibitem{Kronfeld:1991kp}A.~S.~Kronfeld and B.~Nizic,   Phys.~Rev.~D \textbf{44} (1991) 3445 [Erratum-ibid.~D \textbf{46}
 (1992) 2272].
 
 \bibitem{Brooks:2000nb}T.~C.~Brooks and L.~J.~Dixon,   Phys.~Rev.~D \textbf{62} (2000) 114021 [arXiv:hep-ph/0004143].
 
 \bibitem{Lepage:1979zb}G.~P.~Lepage and S.~J.~Brodsky,   Phys.~Lett.~B \textbf{87} (1979) 359.
 
 \bibitem{Lepage:1980fj}G.~P.~Lepage and S.~J.~Brodsky,   Phys.~Rev.~D \textbf{22} (1980) 2157.
 
 \bibitem{Diehl:1998kh}M.~Diehl, T.~Feldmann, R.~Jakob and P.~Kroll,
 Eur.~Phys.~J.~C \textbf{8} (1999) 409 [arXiv:hep-ph/9811253].
 
 \bibitem{Diehl:2004cx}M.~Diehl, T.~Feldmann, R.~Jakob and P.~Kroll,
 Eur.~Phys.~J.~C \textbf{39} (2005) 1 [arXiv:hep-ph/0408173].
 
 \bibitem{Klein:2009zz}M.~Klein and P.~Newman, CERN Cour. \textbf{49N3}
 (2009) 22. 
 
 \bibitem{LHeC}LHEC website \url{http://www.ep.ph.bham.ac.uk/exp/LHeC/}
 
 \bibitem{EIC}EIC website \url{http://www.phenix.bnl.gov/WWW/publish/abhay/Home_of_EIC/}
 
 \bibitem{EIC2}EIC White paper \url{http://www.phenix.bnl.gov/WWW/publish/abhay/Home_of_EIC/NSAC2007/070424_EIC.pdf}
 
 \bibitem{Abelev:2007nb}B.~I.~Abelev \emph{et al.} [STAR Collaboration],
 Phys.~Rev.~C \textbf{77} (2008) 034910 [arXiv:0712.3320 [nucl-ex]].
 
 \bibitem{Adams:2004rz}J.~Adams \emph{et al.} [STAR Collaboration],
 Phys.~Rev.~C \textbf{70} (2004) 031902 [arXiv:nucl-ex/0404012].
 
 \bibitem{Adler:2002sc}C.~Adler \emph{et al.} [STAR Collaboration],
 Phys.~Rev.~Lett.~\textbf{89} (2002) 272302 [arXiv:nucl-ex/0206004].
 
 \bibitem{dEnterria:2006ep}D.~G.~d'Enterria,   arXiv:nucl-ex/0601001.
 
 \bibitem{Baltz:2007kq}K.~Hencken \emph{et al.}, Phys.~Rept.~\textbf{458}
 (2008) 1 [arXiv:0706.3356 [nucl-ex]].
 
 \bibitem{Kopeliovich:1981}B.~Z.~Kopeliovich, L.~I.~Lapidus and
 A.~B.~Zamolodchikov, JETP Lett. \textbf{33} (1981) 595 [Pisma
 Zh. Eksp. Teor. Fiz. \textbf{33} (1981) 612].
 
 \bibitem{mueller} A.~H.~Mueller,   Nucl.\ Phys.\ B \textbf{335}, 115 (1990); A.~H.~Mueller and B.~Patel,
 Nucl.\ Phys.\ B \textbf{425}, 471 (1994) [arXiv:hep-ph/9403256].
 
 \bibitem{Dorokhov:2006qm} A.~E.~Dorokhov, W.~Broniowski and E.~Ruiz
 Arriola,   Phys.\ Rev.\ D \textbf{74} (2006) 054023 [arXiv:hep-ph/0607171].
 
 \bibitem{Kopeliovich:2008ct}B.~Z.~Kopeliovich, I.~Schmidt and
 M.~Siddikov, Phys.~Rev.~D \textbf{79} (2009) 034019 [arXiv:0812.3992
 [hep-ph]].
 
 \bibitem{Bronzan:1974jh}J.~B.~Bronzan, G.~L.~Kane and U.~P.~Sukhatme,
 Phys.~Lett.~B \textbf{49} (1974) 272.
 
 \bibitem{Schafer:1996wv}T.~Schafer and E.~V.~Shuryak, Rev.~Mod.~Phys.~\textbf{70}
 (1998) 323 [arXiv:hep-ph/9610451].
 
 \bibitem{Diakonov:1985eg}D.~Diakonov and V.~Y.~Petrov, Nucl. Phys.
 B \textbf{272} (1986) 457
 
 \bibitem{Diakonov:1995qy} D.~Diakonov, M.~V.~Polyakov and C.~Weiss,
 Nucl.~Phys.~B~ \textbf{461} (1996) 539 [arXiv:hep-ph/9510232].
 
 \bibitem{Anikin:2000rq}I.~V.~Anikin, A.~E.~Dorokhov and L.~Tomio,
 Phys.~Part.~Nucl. \textbf{31} (2000) 509 [Fiz.~Elem.~Chast.~Atom.~Yadra
 \textbf{31} (2000) 1023].
 
 \bibitem{Dorokhov:2003kf}A.~E.~Dorokhov and W.~Broniowski, Eur.~Phys.~J.~C
 \textbf{32} (2003) 79 [arXiv:hep-ph/0305037].
 
 \bibitem{Goeke:2007j}K.~Goeke, M.~M.~Musakhanov and M.~Siddikov,
 Phys.~Rev.~D \textbf{76} (2007) 076007 [arXiv:0707.1997 [hep-ph]]
 
 \bibitem{kst2} B.Z.~Kopeliovich, A.~Sch�fer and A.V.~Tarasov,
 Phys. Rev. D\textbf{62}, 054022 (2000).
 
 \bibitem{gbw} K.~J.~Golec-Biernat and M.~Wusthoff,  Phys.~Rev.~D \textbf{59} (1999) 014017 [arXiv:hep-ph/9807513].
 
 \bibitem{Kopeliovich:2007fv} B.~Z.~Kopeliovich, H.~J.~Pirner,
 A.~H.~Rezaeian and I.~Schmidt,   Phys.\ Rev.\ D \textbf{77} (2008) 034011 [arXiv:0711.3010 [hep-ph]].
 
 \bibitem{Kopeliovich:2008nx} B.~Z.~Kopeliovich, A.~H.~Rezaeian
 and I.~Schmidt,   arXiv:0809.4327 [hep-ph], to appear in Phys. Rev. D.
 
 \bibitem{Kopeliovich:2008da} B.~Z.~Kopeliovich, I.~K.~Potashnikova,
 I.~Schmidt and J.~Soffer,   arXiv:0805.4534 [hep-ph].  
 
 \bibitem{Donnachie:1992ny}A.~Donnachie and P.~V.~Landshoff,   Phys.~Lett.~B \textbf{296} (1992) 227 [arXiv:hep-ph/9209205].
 
 \bibitem{Chekanov:2001gw}S.~Chekanov \emph{et al.} [ZEUS Collaboration],
 Nucl.~Phys.~B \textbf{627} (2002) 3 [arXiv:hep-ex/0202034].
 
 \bibitem{Brandt:1997gi}A.~Brandt \emph{et al.} [UA8 Collaboration],
 Nucl.~Phys.~B \textbf{514} (1998) 3 [arXiv:hep-ex/9710004]. 
 
 \bibitem{Danagoulian:2007gs}A.~Danagoulian \emph{et al.} [Hall
 A Collaboration], Phys.~Rev.~Lett.~\textbf{98} (2007) 152001 [arXiv:nucl-ex/0701068].
 
 \bibitem{Brodsky:1973kr}S.~J.~Brodsky and G.~R.~Farrar,   Phys.~Rev.~Lett. \textbf{31} (1973) 1153.

\bibitem{Aktas:2006hx}
  A.~Aktas {\it et al.}  [H1 Collaboration],
    Eur.\ Phys.\ J.\  C {\bf 48} (2006) 749
  [arXiv:hep-ex/0606003].
  \bibitem{Irving:1976rk}
  A.~C.~Irving,
    Nucl.\ Phys.\  B {\bf 121}, 176 (1977).

\end{thebibliography}
 \end{document}